\documentclass[aps,pra,
a4paper,
amsmath,amssymb,floatfix,twocolumn,superscriptaddress]{revtex4}
\usepackage{geometry}
\usepackage{graphicx}
\usepackage{mathrsfs}
\usepackage{amssymb}
\usepackage{amsmath}
\usepackage{amsthm}
\usepackage{amsbsy}
\usepackage{bm}
\usepackage{hyperref}
\usepackage[T1]{fontenc}

\newtheorem{proposition}{Proposition}
\newtheorem{theorem}{Theorem}

\theoremstyle{definition}

\newtheorem{example}{Example}

\begin{document}

\title{Lower and upper bounds on nonunital qubit channel capacities}

\author{Sergey N. Filippov}

\affiliation{Moscow Institute of Physics and Technology,
Institutskii Per. 9, Dolgoprudny, Moscow Region 141700, Russia}

\affiliation{Institute of Physics and Technology of the Russian
Academy of Sciences, Nakhimovskii Pr. 34, Moscow 117218, Russia}

\begin{abstract}
Classical capacity of unital qubit channels is well known, whereas
that of nonunital qubit channels is not. We find lower and upper
bounds on classical capacity of nonunital qubit channels by using
a recently developed decomposition technique relating nonunital
and unital positive qubit maps.
\end{abstract}

\maketitle

\section{Introduction}

Transmission of classical information through quantum channels is
considered in a series of
papers~\cite{holevo-1977,holevo-1998,schumacher-1997,holevo-rus-1998,king-2001,holevo-2001,holevo-shirokov-2006,shirokov-2006}
and
reviews~\cite{holevo-giovannetti-2012,holevo-2012,wilde-2013,gyongyosi-2018}.
In brief, if a real number $R$ is an achievable rate of
information transmission, then $n$ qubits effectively allow one to
transmit $2^{nR}$ classical messages. The encoder assigns an
$n$-qubit density operator $\varrho_i^{(n)}$ to each message $i$.
The $n$-qubit density operator is a positive semidefinite operator
with unit trace, which acts on a $2^n$ dimensional Hilbert space
${\cal H}_{2^n}$. In the process of information transmission, each
qubit is transmitted through a quantum channel $\Phi$, which is a
completely positive and trace preserving map. Therefore, the
output state of $n$ qubits reads $\Phi^{\otimes n}
[\varrho_i^{(n)}]$. The decoder represents the measurement device
described by a positive operator-valued measure, which assigns a
positive-semidefinite operator $M_j^{(n)}$ (acting on
$2^n$-dimensional Hilbert space) to each observed outcome $j \in
\{1,\ldots,N\}$. Let $p(j|i)$ be the probability of observing
outcome $j\in\{0,1,\ldots,N\}$ if the original message is $i$,
then by the quantum-mechanical rule
\begin{equation} \label{born-rule}
p^{(n)}(j|i) = {\rm tr}[\varrho_i^{(n)} M_j^{(n)}].
\end{equation}

\noindent Condition $\sum_{j=1}^N M_j^{(n)} = I$ guarantees
$\sum_{j=1}^N p^{(n)}(j|i) = 1$. The maximum error probability
reads
\begin{equation}
p_{\rm err}(n,N) = \max_{j=1,\ldots,N} \Big( 1-p^{(n)}(j|j) \Big).
\end{equation}

\noindent $R$ is called an achievable rate of information
transmission if
\begin{equation}
\lim_{n \rightarrow \infty} p_{\rm err}(n,2^{nR}) = 0.
\end{equation}

\noindent By classical capacity $C(\Phi)$ of quantum channel
$\Phi$ we understand the supremum of achievable rates:
\begin{equation}
C(\Phi) = \sup \left\{ R :  \lim_{n \rightarrow \infty} p_{\rm
err}(n,2^{nR}) = 0 \right\}.
\end{equation}

\noindent The Holevo--Schumacher--Westmoreland
theorem~\cite{holevo-1998,schumacher-1997} states that
\begin{equation}
\label{capacity} C(\Phi) = \lim_{n \rightarrow \infty} \frac{1}{n}
C_{\chi}(\Phi^{\otimes n}),
\end{equation}

\noindent where the quantity $C_{\chi}(\Phi)$ is expressed through
all possible ensembles of density operators $\{p_k,\rho_k\}$ and
von Neumann entropy $S(\rho) = - {\rm tr}(\rho {\rm log}_2 \rho)$
via formula
\begin{equation}
\label{chi-capacity} C_{\chi}(\Phi) = \sup_{\{p_k,\rho_k\}} \left[
S\left( \sum_k p_k \Phi[\rho_k] \right) - \sum_k p_k
S(\Phi[\rho_k]) \right].
\end{equation}

\noindent The quantity $C_{\chi}(\Phi)$ is also known as
$\chi$-capacity and the Holevo capacity of quantum channel $\Phi$.

If in formula~(\ref{born-rule}) one restricts to product input
states $\varrho_i^{(n)} = \bigotimes_{k=1}^n \varrho_k^{(1)}(i)$
and separable measurements $M_j^{(n)} = \sum_{x: \ g(x)=j}
\bigotimes_{k=1}^n M_{x_k}^{(1)}$ with a classical data processing
$g: \ x=(x_1,\ldots,x_n) \mapsto j$, then one obtains a modified
capacity $C^{(1)}(\Phi)$. By construction $C^{(1)}(\Phi) \leq
C_{\chi}(\Phi)$. There exists a class of so-called pseudoclassical
channels~\cite{fujiwara-1998} for which $C^{(1)}(\Phi) =
C_{\chi}(\Phi)$. Such a class includes all unital qubit channels
and some nonunital qubit channels too. A complete characterization
of pseudoclassical qubit channels is given in the
paper~\cite{fujiwara-1998}. In this paper, we consider general
nonunital qubit channels without restriction to the class
pseudoclassical ones.

Calculation of the $\chi$-capacity $C_{\chi}(\Phi)$ is complicated
even for the case of a general qubit
channel~\cite{nagaoka-1998,knr-2002,hayashi-2005,daems-2009,renner-2016}
and no closed formula is known. Needless to say, the regularized
capacity $C(\Phi)$ is even more difficult to estimate due to the
fact that additivity hypothesis is not proved for a general qubit
channel. In this paper, we partially fill the gap and find lower
and upper bounds on the capacity $C(\Phi)$ of qubit channels
$\Phi$. We also compare these bounds with the known ones that can
be computed with the help of semidefinite
programming~\cite{wang-2018,leditzky-2018}. We demonstrate that
for some channels our upper bound outperforms the previously known
upper bounds.

\section{Relation between unital and nonunital qubit channels}

Let $A$ and $B$ be two operators acting on $\mathcal{H}_2$. By
$\Phi_A$ we denote a completely positive map $\Phi_A[X] = A X
A^{\dag}$, i.e. a map with a single Kraus operator $A$.
Analogously, $\Phi_B[X] = B X B^{\dag}$. Hereafter, $\dag$ denotes
Hermitian conjugation.

Suppose that $\Phi$ is a qubit linear map, which belongs to the
interior of the cone of positivity preserving maps. Such maps are
also referred to as positivity improving ones~\cite{georgiou-2015}
because $\Phi[\varrho] > 0$ for all $\varrho \geq 0$, $\varrho
\neq 0$. In terms of the paper~\cite{gurvits-2004}, $\inf\{{\rm
det} \Phi[X] \, | \, X > 0, {\rm det}X = 1 \}$ is strictly
positive and attained. Equivalently, $\Phi$ can be represented as
a nontrivial convex combination of some positive map and the
tracing map $X \mapsto {\rm tr}[X] \frac{1}{2} I$, see
Ref.~\cite{aubrun-2015}.

If $\Phi$ belongs to the interior of the cone of positivity
preserving maps, then by Proposition 2.32 in~\cite{aubrun-2017}
there exist positive definite operators $A$ and $B$ acting on
$\mathcal{H}_2$ such that the map
\begin{equation}
\label{Upsilon} \Upsilon = \Phi_A \circ \Phi \circ \Phi_B
\end{equation}

\noindent is trace preserving and unital. Unitality means that
$\Upsilon[I] = I$, where $I$ is the identity operator.

The relation (\ref{Upsilon}) is a quantum analogue of Sinkhorn's
theorem for square matrices with strictly positive
elements~\cite{sinkhorn-1964}. The Sinkhorn theorem states that if
$X$ is an $n \times n$ matrix with strictly positive elements,
then there exist diagonal matrices $D_1$ and $D_2$ with strictly
positive diagonal elements such that $Y = D_1 X D_2$ is doubly
stochastic. In quantum case, $X$ is replaced by $\Phi$, $Y$ is
replaced by $\Upsilon$, $D_1$ and $D_2$ are replaced by $\Phi_A$
and $\Phi_B$, respectively. This is the reason why (\ref{Upsilon})
is sometimes referred to as Sinkhorn's normal form for positive
maps~\cite{aubrun-2017}. Historically, the relation
(\ref{Upsilon}) was originally observed in~\cite{gurvits-2004},
rediscovered for positivity improving completely positive maps
$\Phi$ in~\cite{georgiou-2015}, and finally clarified
in~\cite{aubrun-2015,aubrun-2017}.

Suppose that in addition to being positivity improving $\Phi$ is
also completely positive and trace preserving, then $\Upsilon$ is
completely positive and trace preserving too. For a given
nonunital qubit channel $\Phi$ the particular form of operators
$A$ and $B$ is derived in Refs.~\cite{fm-2017,ffk-2017}. Since $A$
and $B$ are nondegenerate, formula (\ref{Upsilon}) implies that
\begin{equation}
\label{decomposition} \Phi = \Phi_{A^{-1}} \circ \Upsilon \circ
\Phi_{B^{-1}},
\end{equation}

\noindent i.e. all non-boundary nonunital qubit channels $\Phi$
can be decomposed into a concatenation of three completely
positive maps $\Phi_{B^{-1}}$, $\Upsilon$, $\Phi_{A^{-1}}$, with
$\Upsilon$ being unital.

On the other hand, for any unital qubit channel $\Upsilon$ there
exist unitary operators $V$ and $W$ such that~\cite{ruskai-2002}
\begin{equation}
\Upsilon = \Phi_W \circ \Lambda \circ \Phi_V,
\end{equation}

\noindent where the quantum channel $\Lambda$ has so called
diagonal form in the basis of conventional Pauli operators
$I,\sigma_1,\sigma_2,\sigma_3$:
\begin{equation}
\label{Lambda} \Lambda[X] = \frac{1}{2} {\rm tr}[X] I +
\frac{1}{2} \sum_{i=1}^3 \lambda_i {\rm tr}[\sigma_i X] \sigma_i.
\end{equation}

\noindent Parameters $\lambda_1,\lambda_2,\lambda_3$ in
(\ref{Lambda}) are real and satisfy the constraint $1 \pm
\lambda_3 \geq | \lambda_1 \pm \lambda_2 |$ as $\Lambda$ is
completely positive~\cite{ruskai-2002}.

Clearly, classical capacities of channels $\Upsilon$ and $\Lambda$
coincide. Moreover, since the additivity hypothesis holds true for
unital qubit channels~\cite{king-2002}, the classical capacity
equals Holevo capacity and reads
\begin{equation}
\label{C-Upsilon} C(\Upsilon) = C(\Lambda) = C_{\chi}(\Lambda) = 1
- h \left( \frac{1}{2} \Big( 1-\max_{i=1,2,3} |\lambda_i| \Big)
\right),
\end{equation}

\noindent where $h(x) = - x {\rm log}_2 x - (1-x) {\rm log}_2
(1-x)$.

In what follows, we relate the classical capacity of nonunital
qubit channel $\Phi$ with the classical capacity of unital qubit
channel $\Upsilon$, which is given by formula~(\ref{C-Upsilon}).

\section{Bounds on classical capacity of nonunital qubit channels}

\begin{theorem}{Theorem} \label{theorem}
Suppose $\Phi$ is a qubit channel such that the map $\Psi = \Phi_A
\circ \Phi \circ \Phi_B$ is a channel (completely positive and
trace preserving). Then $C(\Phi) \geq C(\Psi) - 2 \log_2 (\|A\|
\|B\|)$.
\end{theorem}

{\it Proof}. Let $\{\varrho_i^{(n)},M_i^{(n)}\}_{i=1}^N$ be the
optimal code of size $N = 2^{n R_{\Psi}}$ for the composite
channel $\Psi^{\otimes n}$ such that $\lim_{n \rightarrow \infty}
p_{{\rm err} \ \Psi}(n,2^{nR_{\Psi}}) = 0$.

Consider a set of modified input states
\begin{equation} \label{modifiedcode}
\widetilde{\varrho}_i^{(n)} = \frac{B^{\otimes n} \varrho_i^{(n)}
(B^{\dag})^{\otimes n}}{{\rm tr}[B^{\otimes n} \varrho_i^{(n)}
(B^{\dag})^{\otimes n}]}.
\end{equation}

\noindent and a modified positive operator-valued measure $\{j
\rightarrow \widetilde{M}_j^{(n)} \}_{j=0}^{N}$ with elements
\begin{eqnarray}
&& \widetilde{M}_0^{(n)} = I - \sum_{j=1}^{N}
\widetilde{M}_j^{(n)},
\\
&& \widetilde{M}_j^{(n)} = \frac{(A^{\dag})^{\otimes n} M_j^{(n)}
A^{\otimes n}}{\|A\|^{2n}}, \quad j=1,\ldots,N,
\end{eqnarray}

\noindent where $\|X\| = \|X\|_{\infty} = \max_{\psi:
\langle\psi|\psi\rangle = 1} \langle\psi| \sqrt{X^{\dag}X}
|\psi\rangle$ is the operator norm. It is not hard to see that
$\widetilde{M}_0^{(n)}$ is positive semidefinite.

Using the modified code, let each qubit be transmitted through the
channel $\Phi$. Then the probability to observe outcome $j \neq 0$
provided input message $i$ equals
\begin{eqnarray}
&& \widetilde{p}^{(n)}(j|i) = {\rm tr}\left[
\widetilde{\varrho}_i^{(n)} \widetilde{M}_j^{(n)} \right]
\nonumber\\
&& = \frac{{\rm tr}\left\{ A^{\otimes n} \Phi^{\otimes n} \left[
B^{\otimes n} \varrho_i^{(n)} (B^{\dag})^{\otimes n} \right]
(A^{\dag})^{\otimes n} M_j^{(n)} \right\} }{ {\rm tr}[B^{\otimes
n} \varrho_i^{(n)} (B^{\dag})^{\otimes n}] \|A\|^{2n}}. \quad
\end{eqnarray}

Since $\Phi_A \circ \Phi \circ \Phi_B = \Psi$, we get
\begin{eqnarray}
\widetilde{p}^{(n)}(j|i) &=& \frac{{\rm tr}\left\{ \Psi^{\otimes
n}[\varrho_i^{(n)}] M_j^{(n)} \right\}}{{\rm tr}[B^{\otimes n}
\varrho_i^{(n)} (B^{\dag})^{\otimes n}] \|A\|^{2n}} \nonumber\\
&=& \frac{p^{(n)}(j|i)}{{\rm tr}[B^{\otimes n} \varrho_i^{(n)}
(B^{\dag})^{\otimes n}] \|A\|^{2n}},
\end{eqnarray}

\noindent where $p^{(n)}(j|i)$ is the probability to get outcome
$j\in\{1,\ldots,N\}$ for the input message $i\in\{1,\ldots,N\}$ in
the original optimal protocol for channel $\Psi^{\otimes n}$.

Observation of the outcome $j=0$ in the modified protocol would be
treated as unsuccessful event, whereas observation of the outcome
$j\in\{1,\ldots,N\}$ leads to a successful identification of the
message because $p^{(n)}(j|i) \rightarrow \delta_{ij}$ if $n
\rightarrow \infty$.

The probability to observe nonzero outcome $j$ equals
\begin{eqnarray}
P^{(n)} = \sum_{j=1}^N \widetilde{p}^{(n)}(j|i) & = &
\frac{1}{{\rm tr}[B^{\otimes n} \varrho_i^{(n)}
(B^{\dag})^{\otimes n}] \|A\|^{2n}} \nonumber\\ & \geq &
\frac{1}{\left( \|A\| \|B\| \right)^{2n}}.
\end{eqnarray}

Utilizing the modified protocol, one can transmit information only
in the case of successful events $j \neq 0$, so the average number
of successfully transmitted messages $\widetilde{N}$ equals
\begin{equation}
\widetilde{N} = P^{(n)} N =  P^{(n)} 2^{n R_{\Psi}} \geq 2^{n
\left(R_{\Psi} - 2 \log_2 (\|A\| \|B\|) \right)}.
\end{equation}

Therefore, the considered protocol enables one to achieve the rate
\begin{equation}
\label{achievable-rate} \widetilde{R} \geq R_{\Psi} - 2 \log_2
(\|A\| \|B\|)
\end{equation}

\noindent by utilizing the channel $\Phi$.

If $R_{\Psi} \leq C(\Psi)$ and one observes the successful event
($j \neq 0$), than the maximum error probability in the modified
protocol
\begin{eqnarray}
&& \widetilde{p}_{\rm err}(n,\widetilde{N}) = \max_{j=1,\ldots,N}
\left( 1-\frac{\widetilde{p}^{(n)}(j|j)}{P^{(n)}} \right)
\nonumber\\
&& = \max_{j=1,\ldots,N} \left( 1-p^{(n)}(j|j) \right) \rightarrow
0 \quad {\rm if} \quad n \rightarrow \infty.
\end{eqnarray}

Taking supremum on both sides of Eq.~(\ref{achievable-rate}) with
requirement $\lim_{n \rightarrow \infty}\widetilde{p}_{\rm
err}(n,\widetilde{N})=0$, we get
\begin{equation}
C(\Phi) \geq C(\Psi) - 2 \log_2 (\|A\| \|B\|).
\end{equation}
\rule{5pt}{5pt}

Theorem~\ref{theorem} can be applied to two equalities relating
nonunital and unital qubit channels: $\Upsilon = \Phi_A \circ \Phi
\circ \Phi_B$ and $\Phi = \Phi_{A^{-1}} \circ \Upsilon \circ
\Phi_{B^{-1}}$. As a result, we immediately get upper and lower
bounds on capacity $C(\Phi)$.

\begin{proposition}{Proposition} \label{proposition}
Let $\Phi$ be a unital qubit channel belonging to the interior of
positive qubit maps, then there exist positive definite operators
$A$ and $B$ acting on $\mathcal{H}_2$ such that the map $\Upsilon
= \Phi_A \circ \Phi \circ \Phi_B$ is unital and
\begin{equation} \label{bounds}
- 2 \log_2 (\|A\| \|B\|) \leq C(\Phi) - C(\Upsilon) \leq 2 \log_2
(\|A^{-1}\| \|B^{-1}\|).
\end{equation}
\end{proposition}

{\it Proof}. The statement straightforwardly follows from the
decomposition existence~\cite{aubrun-2017} and
Theorem~\ref{theorem}. \rule{5pt}{5pt}

To illustrate the obtained results, we consider the following
example dealing with 4-parameter nonunital qubit channels.

\begin{example}{Example} \label{example-1}
Consider a nonunital qubit channel of the form
\begin{equation}
\label{phi-nonunital} \Phi[X] = \frac{1}{2} \left( {\rm tr}[X] (I
+ t_3 \sigma_3) + \sum_{j=1}^{3}\lambda_{j} {\rm tr}[\sigma_{j}
\varrho] \sigma_{j} \right),
\end{equation}

\noindent where $t_3$ and $\lambda_1,\lambda_2,\lambda_3$ are real
parameters , which in addition to the condition of complete
positivity also satisfy the inequality $|t_3|+|\lambda_3| < 1$. It
guarantees that $\Phi$ is an interior point of the cone of
positive qubit maps. Ref.~\cite{ffk-2017} provides the explicit
form of decomposition $\Phi = \Phi_{A^{-1}} \circ \Upsilon \circ
\Phi_{B^{-1}}$. We further simplify it and obtain
\begin{eqnarray}
\label{A-case} && \!\!\!\!\!\!\!\!\!\! A = {\rm diag} \left(
\sqrt[4]{(1-t_3)^2-\lambda_3^2} \, , \
\sqrt[4]{(1+t_3)^2-\lambda_3^2} \right), \!\!\!\!\!\!\!\!\!\! \\
\label{B-case} && \!\!\!\!\!\!\!\!\!\! B = \frac{\sqrt{2} \left( 4
\!-\! \left( \sqrt[4]{(1 \!-\! t_3)^2 \!-\! \lambda_3^2} \!-\!
\sqrt[4]{(1 \!+\! t_3)^2 \!-\! \lambda_3^2} \right)^2
\right)^{\!-1/2}}{\sqrt[4]{(1-t_3)^2-\lambda_3^2}
\sqrt[4]{(1+t_3)^2-\lambda_3^2}} \nonumber\\
&& \!\!\!\!\!\!\!\!\!\! \qquad \times {\rm diag} (\sqrt{b_{1}}, \sqrt{b_{2}}), \!\!\!\!\!\!\!\!\!\! \\
&& \!\!\!\!\!\!\!\!\!\! b_{1} \!=\! (1 \!+\! t_3 \!-\! \lambda_3)
\sqrt[4]{(1 \!-\! t_3)^2 \!-\! \lambda_3^2} \!+\! (1 \!-\! t_3
\!+\! \lambda_3)
\sqrt[4]{(1 \!+\! t_3)^2 \!-\! \lambda_3^2}, \nonumber\\
&& \!\!\!\!\!\!\!\!\!\! b_{2} \!=\! (1 \!+\! t_3 \!+\! \lambda_3)
\sqrt[4]{(1 \!-\! t_3)^2 \!-\! \lambda_3^2} \!+\! (1 \!-\! t_3
\!-\! \lambda_3) \sqrt[4]{(1 \!+\! t_3)^2 \!-\! \lambda_3^2}.
\nonumber
\end{eqnarray}

\noindent The unital qubit map $\Upsilon = \widetilde{\Lambda}$
has the form (\ref{Lambda}) with parameters
\begin{eqnarray}
\label{lambda-tilde-1} \widetilde{\lambda}_1 & = &
\frac{2\lambda_1}{\sqrt{(1+\lambda_3)^2-t_3^2}+\sqrt{(1-\lambda_3)^2-t_3^2}},\\
\widetilde{\lambda}_2 & = &
\frac{2\lambda_2}{\sqrt{(1+\lambda_3)^2-t_3^2}+\sqrt{(1-\lambda_3)^2-t_3^2}},\\
\label{lambda-tilde-3} \widetilde{\lambda}_3 & = &
\frac{4\lambda_3}{\left(\sqrt{(1+\lambda_3)^2-t_3^2}+\sqrt{(1-\lambda_3)^2-t_3^2}\right)^2}.
\end{eqnarray}

\noindent Since both $A$ and $B$ are diagonal, the norms $\|A\| =
\max(A_{11}, A_{22})$, $\|B\| = \max(B_{11},B_{22})$, $\|A^{-1}\|
= 1 / \min (A_{11}, A_{22})$, and $\|B^{-1}\| = 1 / \min (B_{11},
B_{22})$. Substituting these norms in
Proposition~\ref{proposition}, we find lower and upper bounds on
capacity $C(\Phi)$. Namely,
\begin{eqnarray}
\label{lower} C(\Phi) & \!\! \geq \!\! & 1 \!-\! h \left(
\frac{1}{2} \Big( 1\!-\!\max_{i=1,2,3}
|\widetilde{\lambda_i}| \Big) \right) \!-\! 2 \log_2 (\|A\| \|B\|), \\
\label{upper} C(\Phi) & \!\! \leq \!\! & 1 \!-\! h \left(
\frac{1}{2} \Big( 1 \!-\! \max_{i=1,2,3} |\widetilde{\lambda}_i|
\Big) \right) \!+\! 2 \log_2 (\|A^{-1}\| \|B^{-1}\|), \nonumber\\
\end{eqnarray}

\noindent where $\widetilde{\lambda}_{i}$, $i=1,2,3$ are given by
formulas (\ref{lambda-tilde-1})--(\ref{lambda-tilde-3}).
\rule{5pt}{5pt}
\end{example}

\begin{example}{Example} \label{example-2}
Let us consider a class of generalized amplitude damping (GAD)
qubit channels as a partial case of Example~\ref{example-1}. The
GAD channel describes the process of qubit dynamics when it
exchanges excitations with the thermal environment at finite
temperature~\cite{nielsen-2000}. In this case, parameters of the
channel (\ref{phi-nonunital}) depend on time $t \geq 0$ as
follows:
\begin{equation}
\lambda_1 = \lambda_2 = e^{- \gamma t}, \quad \lambda_3 = e^{-2
\gamma t}, \quad t_3 = (2 p - 1)(1 - e^{-2 \gamma t}),
\end{equation}

\noindent where $\gamma$ is the energy dissipation rate and ${\rm
diag}(p,1-p)$ is the equilibrium density operator ($0 \leq p \leq
\frac{1}{2}$ is the population of the excited state in thermal
equilibrium with the environment). The direct calculation yields
\begin{eqnarray}
\label{eigenvalues-reduced-gad} && \widetilde{\lambda}_1 =
\widetilde{\lambda}_2  =  \frac{ e^{-\gamma t} }{ f(p,\gamma t) },
\quad \widetilde{\lambda}_3 = \widetilde{\lambda}_1^2 =
\widetilde{\lambda}_2^2, \\
&& \| A \| \| B \| = \sqrt[4]{\frac{1-p}{p}}
\frac{1}{\sqrt{f(p,\gamma t)}}, \\
&& \| A^{-1} \| \| B^{-1} \| = \sqrt[4]{\frac{1-p}{p}} \,
\sqrt{f(p,\gamma t)}, \label{AB-norms}
\\
&& \label{f-function} f(p,\gamma t) =  \sqrt{p(1-p)}
(1-e^{-2\gamma t}) \nonumber\\
&& \qquad + \sqrt{1-p + p e^{-2\gamma t}} \sqrt{p+(1-p)e^{-2\gamma
t}}. \quad
\end{eqnarray}

\noindent Substituting
(\ref{eigenvalues-reduced-gad})--(\ref{AB-norms}) in
(\ref{lower})--(\ref{upper}), we get the following lower and upper
bounds:
\begin{eqnarray}
\label{lower-GAD} C(\Phi_{\rm GAD}) & \geq & C_{\rm
GAD}^{\downarrow} = 1 - h \left( \frac{1}{2} \Big(
1-\frac{e^{-\gamma t}}{f(p,\gamma t)} \Big) \right) \nonumber\\
&& + \log_2
f(p,\gamma t) - \frac{1}{2} \log_2 \frac{1-p}{p}, \\
\label{upper-GAD} C(\Phi_{\rm GAD}) & \leq & C_{\rm
GAD}^{\uparrow} = 1 - h \left( \frac{1}{2} \Big(
1-\frac{e^{-\gamma t}}{f(p,\gamma t)} \Big) \right) \nonumber\\
&& + \log_2 f(p,\gamma t) + \frac{1}{2} \log_2 \frac{1-p}{p}.
\end{eqnarray}

\begin{figure} \label{figure1}
\begin{center}
\includegraphics[width=8cm]{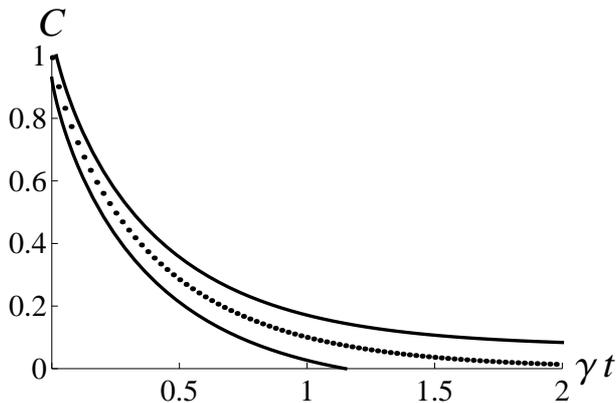}
\caption{Bounds on capacity $C(\Phi_{\rm GAD})$ of the GAD channel
with $p=0.475$ versus dimensionless time $\gamma t$. Lower bound
$C_{\chi}(\Phi_{\rm GAD})$ is plotted in dotted line. The derived
bounds $C_{\rm GAD}^{\downarrow}$ and $C_{\rm GAD}^{\uparrow}$ are
given by lower and upper solid lines, respectively.}
\end{center}
\end{figure}

Figure~1 illustrates these bounds as well as the $\chi$-capacity
of the GAD channel. The latter one can be found numerically since
the structure of optimal ensemble in formula (\ref{chi-capacity})
is known~\cite{berry-2005,li-zhen-2007}. Note that the GAD channel
is not pseudoclassical as it does not satisfy the necessary and
sufficient condition of pseudoclassicality (Theorem~23
in~\cite{fujiwara-1998}) if $p\in(0,\frac{1}{2})$, therefore,
$C^{(1)}(\Phi_{\rm GAD}) < C_{\chi}(\Phi_{\rm GAD}) \leq
C(\Phi_{\rm GAD}) \leq C_{\rm GAD}^{\uparrow}$.

If $p \rightarrow 0$, then the bounds (\ref{lower-GAD}) and
(\ref{upper-GAD}) become trivial. This is due to the fact that the
GAD channel with $p=0$ is a conventional amplitude damping channel
which does not belong to the interior of the cone of positive
maps. Therefore, the products $\|A\|\|B\|$ and $\|A^{-1}\|
\|B^{-1}\|$ diverge if $p \rightarrow 0$. \rule{5pt}{5pt}
\end{example}

\begin{example}{Example} \label{example-3}
Following Ref.~\cite{leditzky-2018}, we consider a one-parameter
qubit channel of the form
\begin{equation} \label{channel-mixture}
\Phi_{\rm mix} = p {\cal A}_p + (1-p) {\cal D}_p,
\end{equation}

\noindent where $0 \leq p \leq 1$, ${\cal A}_p[X] = K_1 X
K_1^{\dag} + K_2 X K_2^{\dag}$ is the qubit amplitude damping
channel with Kraus operators $K_1 = |0\rangle \langle 0| +
\sqrt{1-p} |1\rangle \langle 1|$ and $K_2 = \sqrt{p} |0\rangle
\langle 1|$, ${\cal D}_p$ is the qubit depolarizing channel given
by formula ${\cal D}_p[X] = (1-p) X + \frac{p}{3}(\sigma_x X
\sigma_x + \sigma_y X \sigma_y + \sigma_z X \sigma_z)$.

$\Phi_{\rm mix}$ is a partial case of the 4-parameter channel
discussed in Example~\ref{example-1}:
\begin{eqnarray}
&& \lambda_1 = \lambda_2 = p \sqrt{1-p} + (1-p)
\left(1-\frac{4p}{3}
\right), \\
&& \lambda_3 = (1-p)\left(1-\frac{p}{3} \right), \\
&& t_3 = p^2.
\end{eqnarray}

Substituting these values in equations
(\ref{A-case})--(\ref{lambda-tilde-3}), we readily obtain the
lower and upper bounds, $C_{\rm mix}^{\downarrow}$ and $C_{\rm
mix}^{\uparrow}$, defined by formulas (\ref{lower}) and
(\ref{upper}), respectively. We depict these bounds in Figure~2
and compare them with the previously known lower and upper bounds
for $C(\Phi_{\rm mix})$. One can see that our lower bound is not
as precise as the $\chi$-capacity of $\Phi_{\rm mix}$.
Nevertheless, our upper bound $C_{\rm mix}^{\uparrow}$ is tighter
than the bound in Ref.~\cite{wang-2018} if $p < 0.33$ and tighter
than the bound in Ref.~\cite{leditzky-2018} if $p > 0.29$.
Therefore, for $0.29 < p < 0.33$ our upper bound outperforms the
previously known upper bounds. \rule{5pt}{5pt}
\end{example}

\begin{figure} \label{figure2}
\begin{center}
\includegraphics[width=7.5cm]{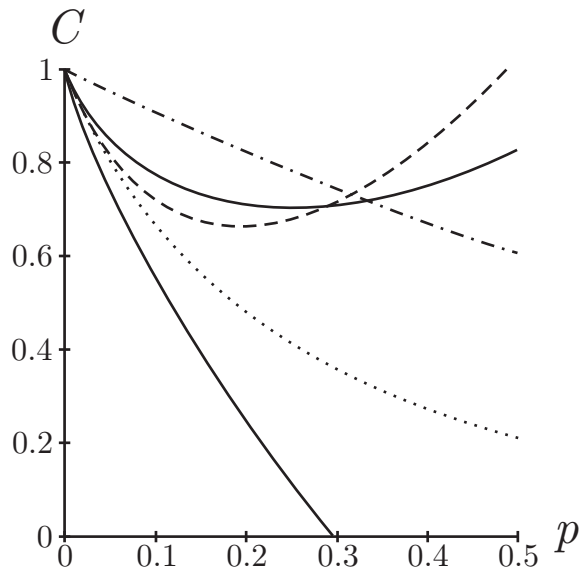}
\caption{Bounds on capacity $C(\Phi_{\rm mix})$ of the mixture
channel (\ref{channel-mixture}) versus dimensionless parameter
$p$. Lower bound $C_{\chi}(\Phi_{\rm mix})$ is plotted in dotted
line. The derived bounds (\ref{lower}) and (\ref{upper}) are
depicted as lower and upper solid lines, respectively. The upper
bound from Ref.~\cite{wang-2018} is plotted in dash-dotted line,
the upper bound from Ref.~\cite{leditzky-2018} is plotted in
dashed line.}
\end{center}
\end{figure}

\section{Conclusions}

We have obtained new lower and upper bounds on classical
capacities of nonunital qubit channels. The obtained result holds
true for the regularized version of Holevo capacity, formula
(\ref{capacity}). Since the optimal coding procedure is known for
unital qubit channels~\cite{king-2002} and relies on the use of
factorized states, our approach is able to provide the factorized
coding (formula (\ref{modifiedcode})), with which one can achieve
the transmission rate corresponding to the lower bound on
capacity.

We illustrate our findings by considering a 4-parameter family of
qubit channels comprising the set of generalized amplitude damping
channels. For some mixtures of amplitude damping and depolarizing
qubit channels the derived upper bound (\ref{upper}) outperforms
the previously known upper bounds of
Refs.~\cite{wang-2018,leditzky-2018}. We believe that the derived
bounds are particularly useful for such channels $\Phi$ that
slightly deviate from unital qubit channels.

Our proofs are based on the seminal relation between unital and
nonunital qubit channels, which was developed in
Ref.~\cite{aubrun-2017}. Such a relation has already been used in
the study of entanglement annihilation~\cite{ffk-2017} and may
turn out to be productive in other research areas as well, for
instance, in the study of absolutely separating quantum
channels~\cite{fmj-2017}, divisibility of qubit dynamical
maps~\cite{wolf-prl-2008,rivas-2010,hall-2014,wudarski-2015,fpmz-2017}
and their tensor products~\cite{benatti-2017}. The most promising
application of the decomposition $\Upsilon = \Phi_A \circ \Phi
\circ \Phi_B$ is in the study of quantum capacities $Q$ of quantum
channels~\cite{lloyd-1997,devetak-2005}. Despite the fact that
both $\Phi_A$ and $\Phi_B$ are completely positive, one cannot
immediately conclude that $Q(\Upsilon)$ is less or equal to
$Q(\Phi)$, neither the inverse statement is justified. The reason
is that the maps $\Phi_A$ and $\Phi_B$ are not trace preserving.
Moreover, at least one of the maps $\Phi_A$ and $\Phi_B$ is not
trace decreasing too. The study of these peculiarities in relation
with quantum capacity and other types of capacities is an
interesting direction for future research.

\begin{acknowledgements}
The study is supported by the Russian Foundation for Basic
Research under Project No. 16-37-60070 mol-a-dk. The author is
grateful to Mark Wilde and Felix Leditzky for useful comments. The
author thanks anonymous referees for helpful suggestions to
improve the quality of the paper.
\end{acknowledgements}

\end{document}